\documentclass [11pt]{article}

\usepackage{axodraw}
\usepackage{graphicx}
\usepackage{amsmath,amssymb} 
\usepackage{latexsym}
\usepackage{mathrsfs}
\usepackage{bbold}
\usepackage{color}
\definecolor{red}{rgb}{1,0,0}

\makeatletter
\def\section{\@startsection {section}{1}{\z@}{-3.5ex plus -1ex minus
 -.2ex}{2.3ex plus .2ex}{\large\bf}}
\def\subsection{\@startsection{subsection}{2}{\z@}{-3.25ex plus -1ex
minus -.2ex}{1.5ex plus .2ex}{\normalsize\bf}}
\makeatother
\makeatletter

\@addtoreset{equation}{section}

\makeatother

\def\bea{\begin{eqnarray}} \def\eea{\end{eqnarray}}
\def\be{\begin{equation}} \def\ee{\end{equation}} \def\nn{\nonumber}
  \def\Z{{\bf Z}}

\newcommand{\promille}{%
  \relax\ifmmode\promillezeichen
        \else\leavevmode\(\mathsurround=0pt\promillezeichen\)\fi}
\newcommand{\promillezeichen}{%
  \kern-.05em%
  \raise.5ex\hbox{\the\scriptfont0 0}%
  \kern-.15em/\kern-.15em%
  \lower.25ex\hbox{\the\scriptfont0 00}}

\begin{document}

\thispagestyle{empty}

\begin{center}

\hfill SISSA-13/2012/EP \\

\begin{center}

\vspace*{0.5cm}

{\Large\bf  On the Cut-off Estimate in Lifshitz}\\[3mm]{\Large\bf  Five Dimensional  Field Theories}

\end{center}

\vspace{1.4cm}

{\bf Marco Serone}\\

\vspace{1.2cm}

{\em SISSA and INFN, Via Bonomea 265, I-34136 Trieste, Italy} 

\vspace{.3cm}

{\em ICTP, Strada Costiera 11, I-34151 Trieste, Italy}

\end{center}

\vspace{0.8cm}

\centerline{\bf Abstract}
\vspace{2 mm}
\begin{quote}

We analyze if and to what extent the high energy behaviour of five-dimensional (5D) gauge theories can be improved by adding certain higher dimensional operators of ``Lifshitz" type, without  breaking  the ordinary four-dimensional Lorentz symmetries.
We show that the UV behaviour of the transverse gauge field polarizations can be improved by the Lifshitz operators, while 
the longitudinal polarizations get strongly coupled at energies lower than the ones in ordinary 5D theories, spoiling the usefulness of the construction in non-abelian gauge theories. We conclude that the improved behaviour as effective theories of the ordinary 5D models is not only related to locality and 5D gauge symmetries, but is a special property of the standard theories defined by the lowest dimensional operators. 

\end{quote}

\vfill

\newpage

\section{Introduction}

Field theories in more than four space-time dimensions have received a lot of attention in recent years. 
They have given us a new perspective on various aspects of high energy physics and cosmology. For instance,  a fundamental TeV-sized quantum gravity scale might arise from extra dimensions  \cite{ArkaniHamed:1998rs} or a TeV scale can naturally emerge from a red-shift effect in a warped extra dimension \cite{Randall:1999ee}.  
Combined with the AdS/CFT idea \cite{Ads-cft}, five-dimensional (5D) theories also give us a new handle to approach strongly coupled QFT \cite{ArkaniHamed:2000ds}.

Field theories in extra dimensions  are non-renormalizable and, as such, they should be seen as effective theories valid up to a maximum energy $\Lambda$, above which they
break down. Estimates based on (not too much) Na\"ive Dimensional Analysis (NDA) and unitarity bounds both give, for a simple 5D gauge theory on a flat segment of length $L=\pi R$,
\be
\Lambda \sim \frac{16 \pi}{g^2R}\,,
\label{Cut-off0}
\ee
where $g$ is the 4D gauge coupling. Based purely on four-dimensional considerations, a 5D gauge theory can be seen as an infinite number of
gauge symmetries,  non-linearly realized by the pseudo Nambu Goldstone Bosons (pNGB's) coming from the gauge field components $A_y^{(n)}$ along the extra dimension $y$, with $m_n=n/R$. 
According to this picture, one would naively expect 
\be
\Lambda_{Naive} \sim 4\pi f
\label{Cut-offNaive}
\ee
where $f = m_1/g$ is the decay constant of the lightest pNGB. We see that $\Lambda = 4/g \Lambda_{Naive}$, and for a sufficiently weak coupling $g$, the 5D theory
remains  weakly coupled up to energies parametrically higher than those expected from a generic 4D effective theory \cite{SekharChivukula:2001hz}. This is also seen in the 4D deconstructed versions of 5D theories \cite{ArkaniHamed:2001ca,ArkaniHamed:2001nc}, where the delay in the unitarity breakdown in scattering amplitudes with respect to the naive estimate arises from non-trivial cancellations among different contributions \cite{Chivukula:2002ej}. 

Aim of this work is to show to what extent the improved high energy behaviour of 5D theories holds and whether it is even possible to improve the situation by modifying the theory by adding higher dimensional operators.
We technically address these questions by analyzing a specific class of non-standard 5D theories with anisotropic scaling symmetry (also called, with some abuse of language, 
Lifshitz field theories,  see \cite{Alexandre:2011kr} for a review and references). The reason to consider these theories is twofold.
First, a symmetry principle allows to restrict the class of higher dimensional operators to consider in studying generalizations of ordinary 5D theories.
Second, Lifshitz field theories are known to possibly have an improved UV behaviour with respect to ordinary theories, exploiting the improved UV behaviour
of the particle propagators. In fact, simple UV-completions of 5D theories based on Lifshitz field theories have been shown to be possible \cite{Iengo:2010xg}.
The price to be paid is however high. In these theories the 4D Lorentz invariance is broken at high energies and is generally recovered in the IR only at the price of extreme fine-tunings \cite{Iengo:2009ix,Collins:2004bp}. For these reasons, we consider here 4D Lorentz-invariant theories, where the anisotropy involves the extra dimension only.\footnote{Five-dimensional Lifshitz theories where the anisotropy of the scaling symmetry shows up only in the extra dimension has been considered in \cite{He:2011hb} and \cite{Hatanaka:2011kg} for a $\lambda \phi^4$ theory  and a model of Gauge-Higgs unification, respectively.} 
We focus on pure non-abelian gauge theories compactified on a plain $S^1/\Z_2$ orbifold, the addition of matter field being straightforward.

We separately study the UV behaviour of the transverse and longitudinal gauge field polarizations.
In the former case we estimate the cut-off by a one-loop computation of the gauge coupling corrections induced by the Kaluza-Klein (KK) modes to the 
zero mode gauge fields.  After showing in some detail the form of this correction in the ordinary 5D case, leading to the cut-off estimate (\ref{NDA0}), we show how the Lifshitz operators
lead to a parametrically higher cut-off, eq.(\ref{NDA3}). The UV behaviour of longitudinal gauge bosons, on the other hand, is analyzed by looking at their elastic scattering amplitudes,
${\cal A}(W_L^{(n)}W_L^{(n)}\rightarrow W_L^{(n)}W_L^{(n)})$, where $n$ is the KK mode of the longitudinal gauge field in the scattering process.
Building on previous results \cite{Csaki:2003dt}, we see that, contrary to the ordinary 5D theories, the ${\cal O}(E^2)$ term in ${\cal A}(W_L^{(n)}W_L^{(n)}\rightarrow W_L^{(n)}W_L^{(n)})$ ($E$ being the center of mass energy), no longer cancels. This leads to the breakdown of unitarity  at energies lower that those obtained in ordinary theories, with an associated cut-off given by eq.(\ref{COLif}), spoiling the usefulness of the Lifshitz construction for non-abelian gauge theories.
In other words, we get that $\Lambda$ can be parametrically higher than the estimate (\ref{Cut-off0}) in abelian gauge theories, while in non-abelian theories the addition of the Lifshitz operators at a sufficiently low scale would result in a decrease of $\Lambda$ to $\Lambda_{Naive}$. 

We conclude that the improved behaviour as effective theories of 5D theories is not only related to locality and 5D gauge symmetries, but is a special property of the standard theories defined by the lowest dimensional operator $F_{MN}^2$. We also deconstruct the simplest version of our Lifshitz 5D theory.  We show that, as expected, the higher dimensional Lifshitz operators are reproduced in the 4D deconstructed model by next-nearest-neighbour terms in field space. 
The precocious breakdown of unitarity induced by the Lifshitz terms is particularly clear from this perspective using the equivalence theorem \cite{Cornwall:1974km}.\footnote{The importance of locality in field space in deconstructed theories has recently been analyzed in \cite{Kahn:2012as}, where it has been shown that non-local terms always lead to a smaller cut-off $\Lambda$. Contrary to the Lifshitz terms considered here, the non-local terms in \cite{Kahn:2012as}  remain non-local in the 5D limit.}

The structure of the paper is as follows. In section 2 we introduce the class of theories we consider. In section 3 we estimate the cut-off $\Lambda$ of these theories by a
one-loop vacuum polarization computation. In section 4 we estimate again the cut-off $\Lambda$, but this time by considering the breakdown of unitarity in scattering amplitudes of longitudinal gauge bosons. In Section 5 we deconstruct a simple 5D Lifshitz theory and show the form of the terms corresponding to the higher dimensional Lifshitz operators.
In section 6 we conclude. We report in Appendix some details of the one-loop vacuum polarization amplitude.


\section{General Set-Up}
\label{sec:scenario}

Lifshitz theories are typically taken to be invariant under anisotropic scale transformations under which the time
coordinate scales differently from the spatial coordinates. In this way higher derivative
terms in the spatial derivatives and quadratic in the fields can be introduced without violation of unitarity.
The improved UV behavior of the propagator turns otherwise non-renormalizable theories in renormalizable ones. 

Along the lines of \cite{He:2011hb}, we consider here Lifshitz models where time and the ordinary spatial directions scale in the same way, so that
this symmetry can be made compatible with the 4D Lorentz symmetry, while the extra dimension scales differently.
We focus on pure 5D non-abelian $SU(m)$ gauge theories (the addition of matter being straightforward) 
compactified on an $S^1/\Z_2$ orbifold of length $L=\pi R$ parametrized by the coordinate $y$,
where the terms odd under the parity symmetry $y\rightarrow - y$ are forbidden and no localized boundary terms are inserted. In this case, the higher derivative Lifshitz terms 
introduced below do not lead to uncancelled boundary terms in the action variation and the 
eigenfunctions of a KK field mode $n$ is the usual $\cos n y/R$ or $\sin n y/R$, depending on the parity symmetry of the field.\footnote{We have not systematically studied the effects of the Lifshitz terms for more general interval compactifications. We expect that new consistency constrains should be imposed 
in this case and more drastic modifications to the spectrum of the theory  might arise.}  
For simplicity, we take in the following  Neumann (Dirichlet) boundary conditions for the gauge fields $A_\mu$ ($A_y$). 
We assume an anisotropic scale invariance of the form
\be
x_\mu = \lambda x_\mu^\prime \,, \ \ \ \ \ y = \lambda^{\frac 1Z} y^{\prime}\,,  \ \ \ \ \
\phi(x^\mu,y) = \lambda^{\frac{Z-d}{2}} \phi^\prime(x^{\mu\prime},y^\prime)\,,
\label{scalingW}
\ee
where $\mu=0,\ldots, 3$ parametrizes the ordinary 3+1 space-time directions, $\phi$ denotes a generic field and $Z$ is a positive integer. 
According to eq.(\ref{scalingW}), we can assign to the coordinates and to the fields a ``weighted'' scaling dimension:
\be
[x^\mu]_w = - 1\,, \ \ \ \  [y]_w = -\frac 1Z\,, \ \ \ \  [\phi]_w = \frac{d-Z}{2}\,.
\label{weightedDim}
\ee
Power-counting renormalizability arguments apply, provided one substitutes the standard scaling dimensions of the operators  by their ``weighted scaling dimensions'' \cite{Anselmi:2007ri}, i.e. by the dimensions implied by the assignment (\ref{weightedDim}).
The weighted dimensions of the gluons $A_\mu$  
are fixed by looking at their ordinary 4D kinetic term components. One gets
\be
[A_\mu]_w = 1+\frac{1}{2Z}\,. 
\ee
Gauge invariance fixes the weighted dimensions of the 5D gauge coupling $g_5$ and of the the gluon components $A_y$:
\be
[g_5]_w = [\partial_\mu]_w - [A_\mu]_w=[\partial_y]_w - [A_y]_w\ \  \ \Longrightarrow \ \ \ [g_5]_w  = -\frac{1}{2Z} \,, \ \ \ [A_y]_w = \frac{3}{2Z}\,.
\ee
For any finite $Z$ the theory remains non-renormalizable, but with a coupling that is less and less irrelevant as $Z$ increases.
Notice that the scaling dimensions of $A_\mu$ and $A_y$ are different, with the latter being smaller than one for $Z>1$.
This difference will play a crucial role in what follows.

The most general Lagrangian involving weighted marginal and relevant operators only (i.e. operators ${\cal O}$ with $[{\cal O}]_w \leq 4+1/Z$) is
\be
{\cal L}=  -\frac 12 {\rm Tr} F_{\mu\nu}^2 +\sum_{i=0}^{Z-1}\frac{a_{i}}{\Lambda_L^{2i}}  {\rm Tr} \, D_y^{i} F_{\mu y}D_y^{i} F_{\mu y}\,, 
\label{L5D}
\ee
where the $SU(m)$ generators $T^a$ in the fundamental representation are normalized
as ${\rm Tr}\, T^a T^b = \delta^{ab}/2$ and $\Lambda_L$ is the energy scale above which the theory effectively behaves as a Lifshitz theory.\footnote{  
It should be clear that, being the theory non-renormalizable, the irrelevant operators we have not written in eq.(\ref{L5D}) cannot be kept to zero at any scale.
When quantum corrections are included, they will be generated.}
By properly rescaling the internal dimension and the scale $\Lambda_L$, without loss of generality, we can set $a_0=a_{Z-1}=1$.
We do not  consider here the problem of understanding where the anisotropic symmetry (\ref{scalingW}) comes from but simply assume its presence in the effective theory. 

The quadratic mixing terms between $A_\mu$ and $A_y$ coming from the second term in eq.(\ref{L5D}) can be canceled by choosing a generalized
$R_\xi$ gauge-fixing term of the form
\be
{\cal L}_{g.f.} = \frac{1}{\xi}{\rm Tr}\, \Big(\partial_\mu A_\mu+\xi \sum_{i=0}^{Z-1}\frac{a_{i}}{\Lambda_L^{2i}} (-1)^i\partial_y^{2i+1} A_y\Big)^2\,.
\label{Lgf}
\ee
The ghost Lagrangian associated to the gauge-fixing (\ref{Lgf}) can easily be derived, though it is not explicitly needed in our analysis. 
The spectrum of states of the Lagrangian (\ref{L5D}) is the standard infinite tower of KK modes labelled by an integer $n$, with the usual 
wave functions of the form
\be
\begin{split}
A_\mu(x,y) = &  \sum_{n=0}^\infty A_\mu^{(n)}(x)\sqrt{\frac{2}{2^{\delta_{n,0}}\pi R}}\ \cos(ny/R)\,, \\
A_y(x,y) = &  \sum_{n=1}^\infty  A_y^{(n)}(x)\sqrt{\frac{2}{\pi R}}\sin(ny/R)\,.
\end{split}
\ee
At the quadratic level, the only effect of the higher derivative Lifshitz terms is to modify the masses of the KK modes:
\be
M_{n}^2  =  \frac{n^2}{R^2} \sum_{i=0}^{Z-1}a_i \frac{n^{2i}}{(\Lambda_L R)^{2i}} \,.
\label{Mgauge}
\ee
The schematic behaviour of the theory is the following. For energies $E<1/R$, it is effectively an ordinary Lorentz invariant 4D gauge theory. For $1/R<E<\Lambda_L$, the theory behaves as an ordinary 5D gauge theory and for $E>\Lambda_L$ it
behaves as a Lifshitz theory where operators are effectively classified by their weighted dimensions. If we take $\Lambda_L \sim 1/R$, the theory is never in the
``ordinary" 5D regime. We will study in the next two sections the impact of the higher derivative Lifshitz operators on the cut-off of the theory, estimated by using gauge coupling corrections and by unitarity bounds on scattering amplitudes.

\section{Estimate of the Cut-off through Gauge Coupling Corrections}

According to NDA, the coefficients of the local operators in a non-renormalizable Lagrangian should be of the same order of the ones induced by radiative corrections at the scale
$\Lambda$, where $\Lambda$ is the energy above which the effective theory breaks down. One can also invert the logic and apply NDA to particularly simple
operators to estimate the value of $\Lambda$ itself. The obvious choice of operator in a 5D gauge theory is the kinetic term $F_{\mu\nu}^2$.
The cut-off $\Lambda$ can then be defined as the scale where the one-loop vacuum polarization correction to the zero mode gauge fields $A_\mu^{(0)}$ becomes of order one.
A naive estimate that just takes into account the phase space of the loop integration and the number of colors would give
\be
\Lambda_{Naive}^{5D} \simeq  \frac{24\pi^3}{m g_5^2} = \frac{24\pi^2}{m g_4^2 R}\,,
\label{Naivelambda}
\ee
where $24\pi^3$ is the 5D loop factor, $g_4=g_5/\sqrt{\pi R}$ is the 4D gauge coupling and $m$ is the quadratic Casimir operator, $C_2(G)=m$, for $SU(m)$. A more detailed computation such as the one below (see appendix for further details) shows that this estimate is in fact too naive and optimistic, and  a more reliable one is obtained by using the 4D loop factor $16\pi^2$ in eq.(\ref{Naivelambda}). In light of these possible discrepancies, in what follows we estimate the cut-off for the Lifshitz field theories by computing in detail the one-loop vacuum polarization for $A_\mu^{(0)}$.

Before considering the Lifshitz case, it is useful to review the ordinary 5D Lorentz invariant computation.
The 5D Lorentz invariant model is obtained by taking $Z=1$ in eq.(\ref{L5D}).
A useful, though not necessary, way to compute the gauge+ghost contribution to the one-loop gauge coupling correction is to make use of a mass-dependent $\beta$-function
in 4D.\footnote{Notice that we use in the following $\beta$-functions and RG flows only as a useful technical tool to get the one-loop correction to the gauge coupling
in 5D. We are not resumming logs.} The whole contribution (ghosts included) of the KK resonances of mass $M_n$ to the $\beta$-function of the 4D gauge coupling  is
(see the appendix D of \cite{Regis:2006hc})
 \be
\beta(g_4,E R)= \frac{g_4^3 }{16\pi^2} \beta_g(E R),
\label{betagauge0}
\ee
with 
\be
\begin{split}
\beta_g(E R) = &  m \bigg(\sum_{n=1}^\infty \int_0^1 \! dx \frac{x(1-x)(6x^2-9x-1) E^2}{M_n^2+E^2 x(1-x)}-\frac{11}{3}\bigg) \\
= &  m \bigg(\sum_{n=-\infty}^\infty \int_0^1 \! dx\frac 12  \frac{x(1-x)(6x^2-9x-1) E^2}{M_n^2+E^2 x(1-x)}-\frac{23}{12}\bigg)\,,
\end{split}
\label{betagauge}
\ee
 where $E$ is the sliding RG (euclidean) energy scale and $M_{n}^2=n^2/R^2$ is the mass of the KK mode $n$.
We show in the appendix some details on how to obtain eq.(\ref{betagauge}), since we are not aware of any derivation in the literature.
The factor $-11/3$ in eq.(\ref{betagauge}) is the zero mode contribution. When $M_n\rightarrow 0$, the integral over $x$ is trivial and gives $-7/2$, which reproduces the contribution
of a massless gauge field plus its scalar (longitudinal) component: $-7/2=-11/3+1/6$.
The one-loop gauge contribution can be written as
\be
g_4^{-2}(E) = g_4^{-2}(E_0) -\frac{1}{8\pi^2}\int_{E_0}^E\frac{ d\mu}{\mu}\, \beta_g(\mu R) \,.
\label{g4Evo}
\ee
Performing the sum over the KK modes $n$, we get 
\be
\beta_g(ER)= m\bigg(  \int_0^1 \!dx \frac{6x^2-9x-1}2\pi ER \sqrt{x(1-x)}
\coth \Big(\pi E  R \sqrt{x(1-x)} \Big)-\frac{23}{12}\bigg)\,, 
\label{betaeffective}
\ee
and, using eq.(\ref{g4Evo}),  the following RG behaviour for $g_4^{-2}$ is obtained:
\be
g_4^{-2}(E) = g_4^{-2}(E_0) -\frac{m}{8\pi^2}\bigg( \int_0^1 \!dx\, \frac{6x^2-9x-1}{2} \log\Big(\frac{\sinh(\pi E  R \sqrt{x(1-x)})}{\sinh( \pi E_0 R \sqrt{x(1-x)})}\Big) -\frac{23}{12}\log \frac{E}{E_0}
\bigg)\,.
\label{RGeffective}
\ee
For $R\rightarrow 0$, eq.(\ref{RGeffective})  reproduces the usual one-loop logarithmic gauge contribution.
We are here interested in using eq.(\ref{betaeffective}) to estimate the cut-off of the theory. The latter is defined as the energy $\Lambda$ where the one-loop factor is comparable to the ``tree-level" term $g^{-2}(E_0)$. For $E\gg 1/R, E_0$, we get
\be
g_4^{-2}(E)\simeq g_4^{-2}(E_0) + \frac{29 m}{1024} E R\,,   \label{RGg5d} 
\ee
from which one obtains
\be
\Lambda^{(1)}\simeq \frac{1024}{29 g^2_0 m}\frac 1R\,.
\label{NDA0}
\ee
In eq.(\ref{NDA0}), $g_0\equiv g_4(E_0)$ and we have introduced a superscript $(1)$ to $\Lambda$ to specify that this is the value of the cut-off for the 
ordinary theory with $Z=1$.  In comparing the naive estimate (\ref{Naivelambda}) with the more refined (\ref{NDA0}) we notice that the former is too optimistic by almost one order of magnitude. If we insist in using naive estimates based on loop factors only, we see that a more reliable estimate is obtained by replacing the 5D loop factor $24\pi^3$ with the
4D loop factor $16\pi^2$ in eq.(\ref{Naivelambda}).

Let us now consider the Lifshitz theory. Interestingly enough, the expression (\ref{betagauge}) for the $\beta$-function still holds, provided  
we use the modified mass terms (\ref{Mgauge}) for the KK gluon mode $n$. This is best seen in unitary gauge, $\xi\rightarrow \infty$, in which $A_y=0$
and the Lifshitz interactions boil down to higher derivative quadratic terms for the KK gluons.
For simplicity, we keep the marginal operators only, setting all couplings $a_i$ to zero,  except $a_{Z-1}=1$. For further simplicity, 
let us first take $Z=2$.  Summing eq.(\ref{betagauge}) over the KK modes $n$ gives
\be
\sum_{n=-\infty}^\infty \frac{1}{n^4+a_2^2} = \frac{1}{a_2^{2}} {\rm Re} \,\Big(\pi\sqrt{a_2} e^{i \pi/4} \cot (\pi \sqrt{a_2}e^{i\pi/4}) \Big)\,,
\ee
where $a_2$ is the value for $Z=2$ of the variable $a_Z$ defined for future use as
\be
a_Z^2 = a_Z^2(E) \equiv  x(1-x) (\Lambda_L R)^{2Z}\frac{E^2}{\Lambda_L^2} \,.
\ee
It is straightforward to check that
\bea
 {\rm Re} \,\Big(\pi\sqrt{a_2} e^{i \pi/4} \cot (\pi \sqrt{a_2}e^{i\pi/4}) \Big) &  = &  \frac{\pi \sqrt{a_2}}{\sqrt{2}} \frac{\sinh (\sqrt{2a_2} \pi) + \sin (\sqrt{2a_2} \pi)} {\cosh (\sqrt{2a_2} \pi) - \cos (\sqrt{2a_2} \pi)} \nn \\
& = & E \frac{d}{d E} \log\Big(\cosh (\sqrt{2a_2} \pi) - \cos (\sqrt{2a_2} \pi)\Big)\,.
\eea
Using the above relations, we get  
\be
g_4^{-2}(E) = g_0^{-2} -\frac{m}{8\pi^2}\bigg( \int_0^1 \!dx \frac{6x^2-9x-1}2
\log\left(\frac{\cosh (\sqrt{2a_2}\pi) -\cos (\sqrt{2a_2} \pi) }{\cosh ( \sqrt{2a_{2,0}}\pi) -\cos ( \sqrt{2a_{2,0}}\pi)} \right) -\frac{23}{12}\log\frac{E}{E_0}\bigg)\,, \ \
\label{RGeffectiveLif}
\ee
where $a_{2,0}=a_2(E_0)$. For $E\gg \Lambda_L, 1/R,E_0$, such that $a_2\gg 1$,we have
\be
g_4^{-2}(E) \simeq g_0^{-2} +\kappa_2 (\Lambda_L R) m \sqrt{\frac{E}{\Lambda_L}}\,,
\label{RGeffectiveUV}
\ee
where 
\be
\kappa_2 =\frac{1}{16\pi^2} \frac{25 \sqrt{2\pi}\Gamma(1/4)^2}{84} \simeq \frac{1}{16}
\ee
is a numerical factor.
The transverse gauge fields $A_\mu^{(0)}$  enter in a strongly coupled regime for
\be
\Lambda^{(2)} \simeq  \Lambda_L \left(\frac{1}{g_0^2 m\kappa_2 (\Lambda_L R)}\right)^2\,.
\label{NDA2}
\ee

We can also analyze the asymptotic region $E\gg 1/R, \Lambda_L, E_0$ for an arbitrary, but finite, $Z$.
We have
\be
\sum_{n=-\infty}^\infty \frac{1}{n^{2Z}+a_Z^2} = \frac{1}{Za^{2}} \sum_{l=0}^{Z-1} {\rm Re} \,\Big( \pi a_Z^{1/Z} e^{\frac{i \pi (2l+1)}{2Z}} \cot (\pi a_Z^{1/Z}e^{\frac{i\pi (2l+1)}{2Z}}) \Big)\,.
\label{sumZgeneric}
\ee
For large energies (i.e. large $a_Z$), we also have
\be
 \sum_{l=0}^{Z-1} {\rm Re} \,\Big( \pi a_Z^{1/Z} e^{\frac{i \pi (2l+1)}{2Z}} \cot (\pi a_Z^{1/Z}e^{\frac{i\pi (2l+1)}{2Z}}) \Big) =  \frac{\pi a_Z^{1/Z}}{\sin\pi/(2Z)} \Big(1+{\cal O}(e^{- c\,  a_Z^{1/Z}})\Big)\,,
\label{largeaEQ}
\ee
where $c$ is a positive numerical factor of ${\cal O}(1)$.
Using the above relations, we get
\be
g^{-2}(E) \simeq g_0^{-2} +m\kappa_Z (\Lambda_L R) \Big(\frac{E}{\Lambda_L}\Big)^{1/Z}\,,
\label{RGeffectiveUVGen}
\ee
where
\be
\kappa_Z =\frac{1}{16\pi^2} \frac{(21Z+8) \pi^{3/2} \Gamma(\frac{1}{2Z})}{8\pi^2 2^{4+1/Z} Z^2 \sin(\frac{\pi}{2Z}) \Gamma(\frac12(5+\frac1Z))}\,.
\label{kappaZ}
\ee
For generic $Z$, the would-be cut-off of the theory is estimated to be
\be
\Lambda^{(Z)} \simeq \Lambda_L \left(\frac{1}{g_0^2 m\kappa_Z (\Lambda_L R)}\right)^Z\,.
\label{NDA3}
\ee
For $Z=1,2$, eq.(\ref{NDA3}) reproduces the previous estimates (\ref{NDA0}) and (\ref{NDA2}).
The first numerical values of $\Lambda^{(Z)}$ are
\be
\Lambda^{(1)} \simeq\frac{35}{g_0^2m} \frac{1}{R}\,, \ \ \ \ 
\Lambda^{(2)} \simeq \frac{260}{(g_0^2m)^2} \frac{\Lambda_L}{(\Lambda_L R)^2}\,, \ \ \ \ 
\Lambda^{(3)} \simeq \frac{940}{(g_0^2m)^3} \frac{\Lambda_L}{(\Lambda_L R)^3}\,, \ \ \  \ldots 
\ee
The would-be cut-off of the theory is parametrically high for a sufficiently small 't Hooft coupling $g_0^2 m$, provided  that $\Lambda_L\sim 1/R$.\footnote{Notice that eq.(\ref{NDA3}) does not hold for parametrically large $Z$ --- in which case it would predict that $\Lambda^{(Z)}\rightarrow 0$ for $Z\rightarrow \infty$ ($k_Z \propto Z$ for large $Z$) --- because the limit of taking large energies does not commute with the large $Z$ limit.} 

Sometimes it is useful to consider how many KK modes  $N_{Max}^{(Z)}$ have a mass below the cut-off of the theory. From an effective field theory point of view, these correspond to the states that we are justified to keep in the theory.\footnote{Naive truncations of this kind should be considered with care, because they can lead to a breakdown of the 5D non-linearly realized gauge symmetries.} Interestingly enough, $N_{Max}^{(Z)}$ does not increase in the Lifshitz theory, because the spacing between the KK modes is enlarged for $Z>1$ and compensates for the higher cut-off $\Lambda^{(Z)}$. Using eqs.(\ref{Mgauge}) and (\ref{NDA3}) we have
\be
N_{Max}^{(Z)} \simeq \frac{1}{g_0^2 m k_Z}\,,
\ee
and since $k_Z$ slightly decreases for increasing values of $Z$, the actual number of KK states below the cut-off actually decreases with respect to the ordinary $Z=1$ theory.

When the weighted relevant operators are considered, $a_i\neq 0$, the sums over the KK modes become rather cumbersome and complicated, but a qualitative physical description
can easily be given. The gauge coupling evolution is essentially dictated by the value of $\Lambda_L R$. For $\Lambda_L R \sim 1$, all the terms appearing in eq.(\ref{Mgauge}) are of the same order of magnitude and the marginal coupling $a_{Z-1}$ quickly dominates for $n>1$. In this case, the approximation above is justified and the would-be cut-off of the theory 
is given by eq.(\ref{NDA3}). For $\Lambda_L R\gg 1$, up to KK modes of order $n\sim \Lambda_L R$,  the dominant coupling is the ordinary $a_0$ term, giving rise to the usual coupling behaviour (\ref{RGg5d}). The theory enters in the Lifshitz regime only for $E>\Lambda_L$.  It is then obvious that the Lifshitz operators are significant only in the energy range  $1/R<\Lambda_L<\Lambda^{(1)}$. 

Similar results also apply in the presence of fermions. The anisotropic scaling (\ref{scalingW}) would demand the presence of higher (covariant) derivative interactions
along the internal dimension, that in unitary gauge boil down to a modified KK mass formula for the fermion KK modes, similar to eq.(\ref{Mgauge}). 
The explicit contribution of a fermion to $\beta$ is reported in appendix, eq.(\ref{betafermion}).
The analysis is essentially identical to the one we did for the gauge case. In particular, the gauge coupling correction
still scales as $(\Lambda_L R) (E/\Lambda_L)^{1/Z}$, as in eq.(\ref{RGeffectiveUVGen}).
The results shown here apply then for abelian theories as well, where at one-loop level the matter contribution is the only one.

\section{Cut-off from Unitarity Bounds in Scattering Amplitudes}

In the last section we have shown that the cut-off of Lifshitz field theories, as obtained by a detailed computation of the vacuum polarization correction of the transverse polarizations of the zero mode field $A_\mu^{(0)}$, can be parametrically higher than the one in ordinary theories.
We show here that the estimate (\ref{NDA3}) does not apply in non-abelian gauge theories, since the scattering amplitudes of longitudinal components of the gauge fields break unitarity well before the energy (\ref{NDA3}) is reached and even before the ordinary 4D value (\ref{NDA0}). This result could also be obtained by analyzing the gauge coupling corrections
to the longitudinal components of $A_\mu^{(n)}$, but this computation is rather cumbersome, while we will see how it is straightforward, building on previous works, to get the bounds coming from scattering amplitudes.

Let us briefly review the behaviour of the scattering amplitudes of longitudinal KK gauge bosons in 5D theories, focusing for simplicity to elastic processes \cite{Csaki:2003dt,SekharChivukula:2001hz}. This amplitude could grow as fast as $E^4$, where $E$ is the center of mass energy of the incoming fields.  In \cite{Csaki:2003dt} it has been shown
that the ${\cal O}(E^4)$ and ${\cal O}(E^2)$ terms in the amplitude of longitudinal 5D gauge boson scattering vanish, 
whenever the following relations hold:
\be
\begin{split}
g_{nnnn}^2 & = \sum_k g_{nnk}^2 \,, \\
4 g_{nnnn}^2 M_n^2 & = 3\sum_k g_{nnk}^2 M_k^2\,, 
\end{split}
\label{sumrules}
\ee
where $k$ and $n$ are KK levels, $g_{nnnn}$ is the quartic gauge coupling of KK gauge fields at level $n$, $g_{nnk}$ is the trilinear coupling
of two KK $n$ and one KK $k$ gauge fields, and $M_n^2$ is the mass of the KK $n$ gauge field. In ordinary 5D theories, eqs.(\ref{sumrules}) are satisfied and the amplitude does not grow with the energy, as already claimed in \cite{SekharChivukula:2001hz,Chivukula:2002ej}.
Unitary violation is detected from the ${\cal O}(E^0)$ terms in the amplitude and arises from the multiplicity
of states in a coupled channel analysis  \cite{SekharChivukula:2001hz}. For a $SU(m)$ theory compactified on a segment,
the maximum number of KK states $N_{Max}$ that can enter in a scattering process without leading to a violation of unitarity is given by \cite{SekharChivukula:2001hz}
\be
N_{Max} \simeq\frac{8\pi}{m} \frac{1}{g_0^2}\,,
\ee
leading to a cut-off estimate 
\be
\Lambda\sim M_{N_{Max}} \simeq  \frac{8\pi}{g_0^2 m R} \,,
\label{Emax}
\ee
roughly in agreement with eq.(\ref{NDA0}). 

As we already mentioned, in the unitary gauge $A_y=0$, no new interactions arise from the higher derivative Lifshitz terms, and  the couplings $g_{nnnn}$ and $g_{nnk}$ are the same as in the ordinary 5D theories. The first constraint in eq.(\ref{sumrules}) is then automatically satisfied.
The only effect of the Lifshitz interactions is to modify the gauge boson KK masses as given in eq.(\ref{Mgauge}). 
It is straightforward to check that, due to the modification in the mass formula,  the second relation in eq.(\ref{sumrules}) is no longer satisfied in the Lifshitz case. 
For illustration, let us consider $Z=2$. For plain $S^1/\Z_2$ compactifications, the sum over $k$ in eq.(\ref{sumrules}) reduces to two terms, $k=0$ and $k=2n$, which are the only
two states that can be exchanged in the scattering process, due to the conservation of the 5D momentum, mod $\Z_2$. A simple computation gives
\be
4 g_{nnnn} M_n^2 - 3 g_{nn0} M_0^2 - 3 g_{nn2n} M_{2n}^2 = -\frac{18}{\pi} g_0^2 \Lambda_L^2  \frac{n^4}{(\Lambda_L R)^4} \,.
\ee
Neglecting the ${\cal O}(E^0)$ terms, the $W_L^{(n)}W_L^{(n)}\rightarrow W_L^{(n)}W_L^{(n)}$ scattering goes like
\be
{\cal A}(W_L^{(n)}W_L^{(n)}\rightarrow W_L^{(n)}W_L^{(n)})_{E^2} \sim \frac{g_0^2}{\Big(1+\frac{n^2}{(\Lambda_L R)^2}\Big)^2}\frac{E^2}{\Lambda_L^2}\,.
\label{WlWlscatt}
\ee
While the transverse components of the gauge fields remain weakly coupled for energies above the ordinary bound (\ref{Emax}), the longitudinal components 
show a breakdown of unitarity at energies below eq.(\ref{Emax}).
We get, from eq.(\ref{WlWlscatt}):\footnote{As before, the actual cut-off should be computed by considering inelastic channels as well,  and can be smaller than the estimate (\ref{COLif}).}
\be
\Lambda \sim \frac{4\pi \Lambda_L }{g_0} \,. 
\label{COLif}
\ee
When $\Lambda_L \simeq 1/R$, eq.(\ref{COLif}) is the energy one would expect from 4D considerations for a pNGB with mass $M_1 \simeq 1/R$ and ``pion" decay constant $f=M_1/g_0$, which would give $\Lambda\simeq 4\pi f$, equal to the naive estimate (\ref{Cut-offNaive}). As expected, the ordinary 5D result  (\ref{Emax}) is recovered for $\Lambda_L \rightarrow \infty$, in which case one has to look at the  ${\cal O}(E^0)$ terms. 

A similar result is obtained, by the equivalence theorem, by studying the scattering of the pNGB's in a different gauge, such as Landau or Feynman gauge.
In these gauges, the higher derivative Lifshitz terms give rise to derivative quartic interactions among the pNGB's that reproduce the behaviour (\ref{WlWlscatt}).
We will briefly come back to this point in the next section, when the 4D deconstructed version of the theory is considered.

\section{Deconstructed 4D Model}

It is interesting to analyze the deconstructed version of our set-up.
Let us briefly recall the deconstruction of an ordinary 5D $SU(m)$ pure gauge theory on an interval \cite{ArkaniHamed:2001ca,ArkaniHamed:2001nc}.
The Lagrangian of a linear moose with $N$ sites and $N-1$ link variables $U_i$ is given by  
\be
{\cal L} = -\frac 12 \sum_{i=1}^N {\rm Tr}\, F_{\mu\nu,i}^2 + f^2 \sum_{i=1}^{N-1}{\rm Tr}\, |D_\mu U_i|^2\,,
\label{Ldec1}
\ee
where the $U_i$'s transform as $U_i\rightarrow g_{i+1} U_i g_{i}^\dagger$ under gauge transformations and 
have only ``nearest-neighbour" interactions with the gauge fields
$A_{\mu,i+1}$ and $A_{\mu,i}$.
The covariant derivative is
\be
D_\mu U_i = \partial_\mu U_i - i g A_{\mu, i+1} U_i + i g U_i A_{\mu,i}\,.
\label{DerU}
\ee
For simplicity we have taken in eqs.(\ref{Ldec1}) and (\ref{DerU}) a universal decay constant $f$ and a universal coupling constant $g$.
The gauge group $SU(m)^{N}$ is non-linearly realized, because the link fields are ``$\sigma$-model" fields that can be written as
\be
U_i(x) = e^{\frac{i \pi_i(x)}{f}}\,,
\ee
in terms of would-be Goldstone bosons $\pi_i(x) = \pi_i^a T^a$. In the unitary gauge $\langle U_i\rangle  = 1$, the Lagrangian (\ref{Ldec1}) contains an $N\times N$
mass matrix for the gauge fields of the form
\be
M^2 = g^2 f^2 \left( \begin{array}{cccccc}
 1 & -1 & 0 & \ldots & 0 & 0\\
 -1 & 2 & -1 & \ldots & 0 & 0  \\
 \vdots & \vdots & \vdots & \ddots & \vdots & \vdots \\
 0 & 0 & 0 & \ldots & 2 & -1  \\
 0 & 0 & 0 & \ldots & -1 & 1  \\
\end{array}
\right)\,,
\label{MDec}
\end{equation}
that has eigenvalues 
\be
M_n^2 = 4 g^2 f^2 \sin^2 \frac{\pi n}{2N} \,, \ \ \ n=0,\ldots, N-1\,.
\ee
We can define
\be
L = N a, \ \ \  a= \frac{1}{gf}\,, \ \ p_5 = \frac{\pi n}L\,,
\label{paraDef}
\ee
so that $a$ can be interpreted as the lattice spacing of the interval and $L$ its length.
The 5D gauge coupling $g_5$ is given by $g_5^2 = a g^2$.
For $n\ll N$, we have
\be
M_n^2 = \frac{4}{a^2}\sin^2 \frac{p_5 a}{2} \simeq p_5^2 = \frac{n^2}{R^2}\,,
\label{Mk0}
\ee
where $R=L/\pi$ is the radius of the 5D covering circle of $S^1/\Z_2$. In the unitary gauge, $\pi_1 = \pi_2 = \ldots = \pi_{N-1}=0$ and
$SU(m)^N$ is spontaneously broken to the diagonal subgroup $SU(m)$. 
In the canonical basis, the gauge coupling $g_4$ of the unbroken
gauge fields is $g_4^2 = g^2/N = g_5^2/L$, in agreement with what expected from a 5D theory.

Let us generalize the deconstruction above and include the higher derivative operators appearing in eq.(\ref{L5D}).
For simplicity, we consider only the $Z=2$ case. The higher derivative terms in the extra dimension suggest
that in the deconstructed theory ``next-nearest-neighbour"  interactions should be present. 
The deconstructed Lagrangian can be written as
\be
{\cal L} = -\frac 12 \sum_{i=1}^N {\rm Tr}\, F_{\mu\nu,i}^2 + f^2 \sum_{i=1}^{N-1} {\rm Tr}\, |D_\mu U_i|^2 + \tilde f^2 \sum_{i=1}^N{\rm Tr}\,  |D_\mu U_i-U_{i+1}^\dagger D_\mu U_{i+1} U_i|^2\,,
\label{LdecL}
\ee
where $U_{N+1}=U_1$,  $U_N=0$, $A_{N+1}=A_1$ and $A_{N+2}=A_2$ in the last term, and $\tilde f$ is the ``Lifshitz"  pion decay constant.  
As we will shortly see, the last term in eq.(\ref{LdecL}) corresponds to the ${\rm Tr}\, (D_y F_{\mu y})^2$ term in eq.(\ref{L5D}).
In the unitary gauge $\langle U_i\rangle =1$, all $\pi_i$'s vanish and
$SU(m)^N$ is spontaneously broken to the diagonal subgroup $SU(m)$,  like in the ordinary case reviewed above.
The quadratic terms for the gauge fields coming from the last term in eq.(\ref{LdecL}) give rise to the following $N\times N$
mass matrix:
\be
\widetilde M^2 = g^2 \tilde f^2 \left( \begin{array}{cccccccc}
 2 & -3 & 1 & 0 &  \ldots & 0 & 0 & 0\\
 -3 & 6 & -4 & 1 & \ldots & 0 & 0 & 0  \\
  1 & -4 & 6 & -4 & \ldots & 0 & 0 & 0  \\
 \vdots & \vdots & \vdots & \vdots & \ddots & \vdots & \vdots &\vdots  \\
  0 & 0 & 0 & 0& \ldots  & 6 &-4 & 1   \\
 0 & 0 & 0 & 0& \ldots  & - 4 &6 & -3   \\
 0 & 0 & 0 & 0  & \ldots & 1 & -3 & 2  \\
\end{array}
\right)\,.
\label{MDecL}
\end{equation}
The total mass matrix for the gauge fields is given by
\be
M_{Tot}^2 = M^2 + \widetilde M^2\,,
\label{Mtot}
\ee
with $M^2$ as in (\ref{MDec}). 
By explicit computation, we find that the mass matrix (\ref{MDecL}) has eigenvalues
\be
\tilde M_n^2 = 16 g^2 \tilde f^2 \sin^4 \frac{\pi n}{2N}\,, \ \ \ n=0,\ldots, N-1\,.
\label{MkL}
\ee
Interestingly enough, the matrices (\ref{MDec}) and (\ref{MDecL}) are simultaneously diagonalizable and hence
the total mass eigenvalues are simply given by the sum of eqs.(\ref{Mk0}) and (\ref{MkL}):
\be
M_{Tot,n}^2 = 4 g^2  f^2 \sin^2 \frac{\pi n}{2N}+ 16 g^2 \tilde f^2 \sin^4 \frac{\pi n}{2N} \,, \ \ \ n=0,\ldots, N-1\,.
\ee
The Lifshitz scale $\Lambda_L$ is defined as
\be
a^2 \Lambda_L = \frac{1}{g\tilde f} \,.
\label{paraLif}
\ee
For $n \ll N$, we have
\be
M_n^2 =\frac{4}{a^2}\sin^2 \frac{p_5 a}{2}+ \frac{16}{\Lambda_L^2 a^4}\sin^4 \frac{p_5 a}{2}  \simeq p_5^2 +\frac{p_5^4}{\Lambda_L^2} = \frac{n^2}{R^2}\Big(1+ \frac{n^2}{(\Lambda_L R)^2}\Big)\,,
\ee
which reproduces eq.(\ref{Mgauge}) for $Z=2$. 

The leading behaviour of the amplitude (\ref{WlWlscatt}) can be reproduced in the deconstructed model. 
It is actually easier, by using the equivalence theorem, to look at the $\pi \pi\rightarrow \pi \pi $ scattering. In the ordinary linear moose (\ref{Ldec1}),
the ${\cal O}(E^2)$ term schematically reads
\be
{\cal A}(\pi \pi \rightarrow \pi \pi)_{E^2}\propto \frac{E^2}{f^2} = g^2\Big(\frac{EL}{N}\Big)^2\,,
\label{ApipiOrd}
\ee
where we have used eq.(\ref{paraDef}) in the second equality. 
For simplicity we have omitted in eq.(\ref{ApipiOrd})  gauge and site indices and have written only the structure of the amplitude.
In the limit $N\rightarrow \infty$, the ${\cal O}(E^2)$ term in the amplitude vanishes, in agreement with the 5D expectation.
In the linear ``Lifshitz"  moose (\ref{LdecL}), additional derivative quartic couplings arise and we get an extra term $\Delta {\cal A}$ contributing to the amplitude:\footnote{The Lifshitz term
in eq.(\ref{LdecL}) also contains additional contributions to the kinetic terms of the pions, that have thus to be canonically normalized. The net effect of this normalization is the term in the denominator appearing in eq.(\ref{WlWlscatt}). For simplicity, we neglect these corrections that do not play an important role for our purposes.}
\be
\Delta {\cal A}(\pi \pi \rightarrow \pi \pi)_{E^2}\propto \frac{E^2 \tilde f^2}{f^4} = g^2\Big(\frac{E}{\Lambda_L}\Big)^2\,,
\label{ApipiLif}
\ee
where in the last equality we have used eqs.(\ref{paraDef}) and (\ref{paraLif}). The factors of $N$ now cancel and the ${\cal O}(E^2)$ term no longer 
vanishes in the 5D limit. Instead, using eq.(\ref{ApipiLif}), we get a cut-off $\Lambda \sim 4\pi \Lambda_L/g$, in agremeent with eq.(\ref{COLif}).


Although we have not explicitly worked out the deconstructed version of eq.(\ref{L5D}) for general $Z$, 
we expect that the introduction of  ``next-next-$\ldots$- nearest-neighbour" interactions should reproduce the corresponding 5D higher-derivative terms for any $Z$.
No new results are expected to arise by considering higher values of $Z$.

\section{Conclusions}

We have studied the cut-off estimate in 5D field theories, where certain higher derivative (Lifshitz) operators are added to the action. 
By a detail one-loop vacuum polarization computation, we have argued that the transverse polarizations of the gauge fields
have a softer UV behaviour with respect to the ones in ordinary 5D theories. 
On the other hand, the same higher derivative terms negatively affect the longitudinal polarizations of the gauge fields.
Because of these operators, the ${\cal O}(E^2)$ terms in the scattering amplitude of longitudinal gauge bosons no longer vanish, in contrast to the usual 5D case,
and lead to an earlier breakdown of unitarity with respect to the standard 5D situation. Of course, this problem does not occur for abelian gauge theories, in which 
the Lifshitz operators do improve the UV behaviour of the theory. We have then considered (for the special case $Z=2$) the deconstructed version of the 5D Lifshitz models
and shown how similar conclusions are reached from this perspective. As expected, the Lifshitz terms correspond to next-nearest-neighbour interactions in field space.

Our analysis explicitly shows that the relatively good UV behaviour of standard 5D theories,
for which $\Lambda>\Lambda_{Naive}$, as defined in eqs.(\ref{Cut-off0}) and (\ref{Cut-offNaive}), do not only come from 5D locality and 5D gauge symmetries, both preserved
in our Lifshitz construction, but are peculiar of the standard 5D action. The 4D deconstructed models are useful in this respect, since they show how the Lifshitz terms break
the global symmetries responsible for the good UV behaviour of ordinary 5D theories. 
From the Lifshitz field theory point of view, our results explicitly show that care has to be used in determining the UV behaviour
(e.g. renormalizability) of Lifshitz theories based only on the effective UV dimension of the couplings. It is crucial to also pay attention to the effective dimensions of the fields,
even when they can be gauged away (like the fields $A_y$ in our case), since they can lead to a precocious strong coupling behaviour.

\section*{Acknowledgments}

I thank Roberto Iengo for discussions at the early stages of this project, Michele Redi for a misprint in section 5 and especially Elena Vigiani that pointed out that the argument
about the protection of the deconstructed Lifshitz terms against radiative corrections, appearing in section 5 in previous versions of the paper, was incorrect.


\appendix

\section{Derivation of the 5D $\beta$-function}

The $\beta$-function (\ref{betagauge}) is conveniently computed using background field methods and a background field gauge fixing.
We write the gauge field as $A_{tot} = \bar A+A$, where $\bar A$ is the classical background value and $A$ is the quantum fluctuation.
The 5D Lagrangian reads, including gauge fixing and ghosts, 
\be
{\cal L}_{BFG} = -\frac{1}{4} F_{MN,a}^{tot,2} -\frac{1}{2\xi} (\bar D_M A^M_a)^2 + (\bar D_M\omega_a) (\bar D^M \omega_a- g f_{abc} \omega_b A_M^c)-\frac{1}{4}\delta Z \bar F_{\mu\nu,a}^2  \,,
\label{BFG}
\ee
with $a,b,c$ color indices, $M,N$ 5D indices and
\be
\bar D_M A_{N,a} = \partial_M A_{N,a} + g f_{abc} \bar A_{M,b} A_{N,c} 
\ee
the covariant derivative with respect to the background field only. In eq.(\ref{BFG}), we have explicitly included the counterterm $\delta Z$ for the 4D
background field strength $\bar F_{\mu\nu}(x)$, omitting all the others that do not play any role in the computation.
We choose in the following $\xi=1$ so that all quadratic gauge mixing terms vanish.
As well known, the gauge symmetries of the classical background allow to compute the $\beta$-function directly from the two-point function 
$\langle \bar A_\mu(-p) \bar A_\nu(p) \rangle$. The effective one-loop Lagrangian for the zero mode background $\bar A_\mu(x)$ reads
\be
{\cal L}_{eff} = -\frac 14 Z \bar F_{\mu\nu}^2 +\ldots 
\ee
We choose a standard momentum subtraction renormalization scheme, by demanding that 
\be
Z(p^2=-E^2)=1.
\label{MOM}
\ee
The mass dependent $\beta$-function is given by
\be
\beta(g_4, ER)  = g_4 \frac{d \log Z}{d \log E}\,.
\label{betadef}
\ee
After a lengthy computation, we get the following results for the relevant Feynman graphs, in dimensional regularization:\footnote{Notice that dimensional regularization is typically
used in association with a mass-independent renormalization scheme, such as MS or $\overline{{\rm MS}}$, in which $\mu$, coming from $g\rightarrow g\mu^{\epsilon/2}$,
is the RG scale. In our (unconventional)
use of dimensional regularization with a mass-dependent scheme, $\mu$ becomes irrelevant and the RG scale is identified with the subtraction scale $E$.} 
\bea
&& \mbox{
 \begin{picture}(80,0) (37,43)
    \SetColor{Black}
      \Gluon(19,44)(49,44){3}{4}
    \Gluon(20,45)(50,45){3}{4}
      \Gluon(79,44)(109,44){3}{4}
       \Gluon(80,45)(110,45){3}{4}
    \DashArrowArc(65,45)(15,0,180){3}
      \DashArrowArc(65,45)(15,180,0){3}
    \Vertex(80,45){2}
    \Vertex(50,45){2}
    \Text(-10,40)[lb]{{\Black{$(a)=$}}}
  \end{picture}} =   -iC_2(G)\frac{g_4^2}{(4\pi)^{d/2}}\Gamma\Big(2-\frac d2\Big)\sum_{n=0}^\infty\int_0^1\!dx \Big(\frac{\mu^2}{M_n^2-p^2x(1-x)}\Big)^{\frac{\epsilon}{2}} \nn  \\ && \nn \\ && \nn \\
&& \hspace{2cm}\times \bigg(-\frac{4}{2-d}\Big(M_n^2-p^2x(1-x)\Big) \eta^{\mu\nu}+ (1-2x)^2p^\mu p^\nu \bigg) \,, 
\label{VP1}
\eea
\vskip 3pt
\bea
&& \mbox{
  \begin{picture}(80,0) (37,43)
      \Text(-9,40)[lb]{{\Black{$(b)=$}}}
    \SetColor{Black}
     \Gluon(19,44)(49,44){3}{4}
    \Gluon(20,45)(50,45){3}{4}
       \Gluon(80,45)(110,45){3}{4}
          \Gluon(79,44)(109,44){3}{4}
    \GlueArc(65,45)(15,0,180){3}{6}
      \GlueArc(65,45)(15,180,360){3}{6}
    \Vertex(80,45){2}
    \Vertex(50,45){2}
  \Text(60,12)[lb]{\footnotesize{\Black{$A_\mu$}}}
  \end{picture}} =   -iC_2(G)\frac{g_4^2}{(4\pi)^{d/2}}\Gamma\Big(2-\frac d2\Big)\sum_{n=0}^\infty\int_0^1\!dx \Big(\frac{\mu^2}{M_n^2-p^2x(1-x)}\Big)^{\frac{\epsilon}{2}} \nn  \\ && \nn \\ && \nn \\ &&  \times \bigg(\Big(-4p^2 +\frac{2d}{2-d}\Big(M_n^2-p^2 x(1-x)\Big)\Big) \eta^{\mu\nu}+ \Big(\frac{8-d}2+2dx(1-x)\Big)p^\mu p^\nu \bigg) \,, 
\label{VP2}
\eea
\vskip 3pt
\bea
&& \mbox{
  \begin{picture}(80,0) (37,43)
      \Text(-9,40)[lb]{{\Black{$(c)=$}}}
    \SetColor{Black}
     \Gluon(19,44)(49,44){3}{4}
    \Gluon(20,45)(50,45){3}{4}
       \Gluon(80,45)(110,45){3}{4}
          \Gluon(79,44)(109,44){3}{4}
    \GlueArc(65,45)(15,0,180){3}{6}
      \GlueArc(65,45)(15,180,360){3}{6}
    \Vertex(80,45){2}
    \Vertex(50,45){2}
  \Text(60,12)[lb]{\footnotesize{\Black{$A_y$}}}
  \end{picture}} =   -iC_2(G)\frac{g_4^2}{(4\pi)^{d/2}}\Gamma\Big(2-\frac d2\Big)\sum_{n=1}^\infty\int_0^1\!dx \Big(\frac{\mu^2}{M_n^2-p^2x(1-x)}\Big)^{\frac{\epsilon}{2}} \nn  \\ && \nn \\ && \nn \\ &&   \hspace{1.6cm}\times \bigg(\Big(\frac{2}{2-d}\Big(M_n^2-p^2 x(1-x)\Big)\Big) \eta^{\mu\nu}+ \Big(-\frac{1}{2}+2x(1-x)\Big)p^\mu p^\nu \bigg) \,, 
\label{VP3}
\eea
\vskip 3pt
\bea
&& \mbox{
  \begin{picture}(80,0) (37,43)
      \Text(-9,40)[lb]{{\Black{$(d)=$}}}
    \SetColor{Black}
    \Gluon(19,44)(64,44){3}{7}
    \Gluon(20,45)(65,45){3}{7}
       \Gluon(65,45)(110,45){3}{7}
          \Gluon(64,44)(109,44){3}{7}
    \GlueArc(65,65)(15,-90,270){3}{13}
    \Vertex(65,47){2.4}
    \Text(30,67)[lb]{\footnotesize{\Black{$A_\mu$}}}
  \end{picture}} = -iC_2(G)\frac{g_4^2}{(4\pi)^{d/2}}\Gamma\Big(2-\frac d2\Big)\sum_{n=0}^\infty\int_0^1\!dx \Big(\frac{\mu^2}{M_n^2-p^2x(1-x)}\Big)^{\frac{\epsilon}{2}} \nn  \\  && \nn \\ &&   \hspace{2.3cm }\times \bigg( d\Big((1-x)^2 p^2 -M_n^2\Big)-\frac{d^2}{2-d}(M_n^2-p^2x(1-x)) \bigg) \eta^{\mu\nu} \,,\\ \nn
  \label{VP5}
\eea
\vskip2pt
\bea
&& \mbox{
  \begin{picture}(80,0) (37,43)
    \SetColor{Black}
    \Gluon(19,44)(64,44){3}{7}
          \Text(-9,40)[lb]{{\Black{$(e)=$}}}
    \Gluon(20,45)(65,45){3}{7}
       \Gluon(65,45)(110,45){3}{7}
          \Gluon(64,44)(109,44){3}{7}
    \GlueArc(65,65)(15,-90,270){3}{13}
    \Vertex(65,47){2.4}
       \Text(30,67)[lb]{\footnotesize{\Black{$A_y$}}}
  \end{picture}} =   -iC_2(G)\frac{g_4^2}{(4\pi)^{d/2}}\Gamma\Big(2-\frac d2\Big)\sum_{n=1}^\infty\int_0^1\!dx \Big(\frac{\mu^2}{M_n^2-p^2x(1-x)}\Big)^{\frac{\epsilon}{2}} \nn  \\  && \nn \\ &&  \hspace{3cm }\times \bigg( (1-x)^2 p^2 -M_n^2-\frac{d}{2-d}(M_n^2-p^2x(1-x)) \bigg) \eta^{\mu\nu}\,, \\ \nn
  \label{VP6}
\eea
\vskip2pt
\bea
&& \mbox{
  \begin{picture}(80,0) (37,43)
    \SetColor{Black}
        \Text(-9,40)[lb]{{\Black{$(f)=$}}}
    \Gluon(19,44)(64,44){3}{7}
    \Gluon(20,45)(65,45){3}{7}
       \Gluon(65,45)(110,45){3}{7}
          \Gluon(64,44)(109,44){3}{7}
    \DashArrowArc(65,64)(15,-90,270){3}
    \Vertex(65,47){2.4}
  \end{picture}} =   -iC_2(G)\frac{g_4^2}{(4\pi)^{d/2}}\Gamma\Big(2-\frac d2\Big)\sum_{n=0}^\infty\int_0^1\!dx \Big(\frac{\mu^2}{M_n^2-p^2x(1-x)}\Big)^{\frac{\epsilon}{2}} \nn  \\  && \nn \\ &&  \hspace{2.9cm }\times \bigg( -2\Big((1-x)^2 p^2 -M_n^2\Big)+\frac{2d}{2-d}(M_n^2-p^2x(1-x)) \bigg) \eta^{\mu\nu}\,. \\ \nn
  \label{VP7}
\eea
In the above graphs $M_n=n/R$ is the mass of the KK mode running into the loop, the single wiggly lines represent the gluon fluctuations $A_\mu$ and $A_y$, dashed lines represent the ghost fields and the double wiggly lines represent the
background field $\bar A_\mu$.  Notice the presence of a quartic interaction among ghost and gauge fields in this gauge, leading to the graph $(f)$.
Summing all the contributions, we get
\be
\begin{split}
(a)+(b)+(c)+(d)+(e)+(f) =  &  iC_2(G)\frac{g_4^2}{16\pi^2}(\eta^{\mu\nu} p^2-p^\mu p^\nu) \int_0^1\!dx \bigg( (-2-6x+4x^2) \\
&\hspace{-5cm}  \times \log  \Big(\frac{-p^2x(1-x)}{\mu^2}\Big)+ \sum_{n=1}^\infty (6x^2-9x-1) \log  \Big(\frac{M_n^2-p^2x(1-x)}{\mu^2}\Big)+C \bigg)+{\cal O}(\epsilon)\,,
\end{split}
\label{Zfactor}
\ee
where $C$ is an irrelevant divergent constant. The finite wave function correction $Z$ is determined by the renormalization condition (\ref{MOM}), that fixes the counter-term $\delta Z$. Using eq.(\ref{betadef}), we can finally get eqs.(\ref{betagauge0}) and (\ref{betagauge}).

For completeness, we also report the contribution from a massless 5D fermion in a representation $r$ of $SU(m)$:
\bea
&& \mbox{
 \begin{picture}(80,0) (37,43)
    \SetColor{Black}
      \Gluon(19,44)(49,44){3}{4}
    \Gluon(20,45)(50,45){3}{4}
      \Gluon(79,44)(109,44){3}{4}
       \Gluon(80,45)(110,45){3}{4}
    \ArrowArc(65,45)(15,0,180)
      \ArrowArc(65,45)(15,180,0)
    \Vertex(80,45){2}
    \Vertex(50,45){2}
  \end{picture}} =   -iT(r)\frac{g_4^2}{(4\pi)^{d/2}}\Gamma\Big(2-\frac d2\Big)\sum_{n=0}^\infty \int_0^1\!dx \Big(\frac{\mu^2}{M_n^2-p^2x(1-x)}\Big)^{\frac{\epsilon}{2}} \nn  \\ && \nn \\ 
&&\hspace{4cm} \times 2^{1-\delta_{n,0}}(-4) x(1-x)  \Big(\eta^{\mu\nu}p^2 - p^\mu p^\nu \Big) \,,
\label{VP8}
\eea
that  gives rise to the following contribution to the $\beta$-function:
\be
\beta(g_4,E R) = \frac{g_4^3}{4\pi^2} \sum_{n=-\infty}^\infty  T(r)\int_0^1 \! dx \frac{x^2(1-x)^2 E^2}{M_{n}^2+E^2 x(1-x)}\,.
\label{betafermion}
\ee

\end{document}